# On the Solvability of 2-pair Unicast Networks — A Cut-based Characterization

Kai Cai*, K. B. Letaief† *Fellow, IEEE,* Pingyi Fan‡, and Rongquan Feng§
*Institute of Computing Technology, Chinese Academy of Sciences, Beijing, China, 100190
†Department of Electronic and Computer Engineering,
Hong Kong University of Science and Technology, Hong Kong, China
‡Department of Electronic Engineering, Tsinghua University, Beijing, China, 100084
§LMAM, School of Mathematical Sciences, Peking University, Beijing, China, 100871
Emails: caikai@software.ict.ac.cn; eekhaled@ece.ust.hk; fpy@tsinghua.edu.cn;
fengrq@math.pku.edu.cn

**Abstract**

In this paper, we propose a subnetwork decomposition/combination approach to investigate the single rate 2-pair unicast problem. It is shown that the solvability of a 2-pair unicast problem is completely determined by four specific link subsets, namely, $\mathcal{A}_{1,1}$, $\mathcal{A}_{2,2}$, $\mathcal{A}_{1,2}$ and $\mathcal{A}_{2,1}$ of its underlying network. As a result, an efficient cut-based algorithm to determine the solvability of a 2-pair unicast problem is presented.

## I. INTRODUCTION

It is an important issue to decide the admissible rate region for a multi-source multi-sink communication network in network information theory. The history of the research can be traced back to the earlier work of Elias *et al.* [1], as well as Ford and Fulkerson [2] in 1956. It was shown that the capacity of every one-source one-sink (point-to-point) network can be characterized by its minimum cut (Max-flow Min-cut Theorem). In [3]-[5], Yeung and Zhang presented the inner and outer bounds of the admissible rate region for a distributed source coding system. Based on these works, Ahlswede *et al.* [6] showed that the Max-flow Min-cut capacity can be achieved for multicast networks by using a coding strategy in their seminal work on network coding. Later on, Li *et al.* [7] proved that linear network coding is sufficient to achieve the Max-flow Min-cut capacity for multicast networks.

Unlike the one source networks, for a general multi-source multi-sink network with arbitrary transmission requirements, the Max-flow Min-cut capacity bound can be quite loose. Although some outer and inner bounds [8]-[12], and an entropy characterization [13] have been proposed, the explicit evaluation of the rate region for a general multi-source multi-sink network is very challenging. So many previous studies concentrated on the $k$-pair networks.

The $k$-pair communication problem, which is also known as the multiple unicast sessions, aims at supporting $k$ independent point-to-point communications. Without network coding, i.e., just using pure routing strategy, it is the conventional multi-commodity flow (MCF) problem. For the MCF problem, a fractional achievable rate can be found using linear programming, but it is generally NP-hard to find an integral solution, except for the directed acyclic case, for which there is a polynomial algorithm of using the pebbling game [14], which is of extraordinary complexity. When considering network coding, it is conjectured that there is no more advantage than using fractional routing in undirected networks. This is known as the *undirected k-pair conjecture* [15] and has been verified just for a few classes (see [8], [15] and [16]). In contrast, network coding can provide a significant rate increase in directed $k$-pair networks [15]. Except for the undirected 2-pair networks (and a few other families, see [15] for reference), whose capacity regions can be characterized by the cut condition, it is very difficult to evaluate the exact rate region for a $k$-pair network in general.

In this paper, we propose a subnetwork decomposition/combination approach to investigate the underlying graph structure of the *directed acyclic 2-pair unicast networks*. Our result shows that the solvability of a 2-pair unicast problem is completely determined by four particular link subsets of the underlying network, namely, $\mathcal{A}_{1,1}$, $\mathcal{A}_{2,2}$, $\mathcal{A}_{1,2}$ and $\mathcal{A}_{2,1}$, which can be considered as the most "important" links of the 2-pair network. As a result, we show that a 2-pair unicast problem is solvable if and only if the underlying network contains a copy of one of the four networks shown in Fig.1. Consequently, an efficient cut-based algorithm to determine the solvability of a 2-pair unicast problem is presented. In addition, a new proof that nonlinear network coding is unnecessary for a 2-pair unicast problem is obtained [1].

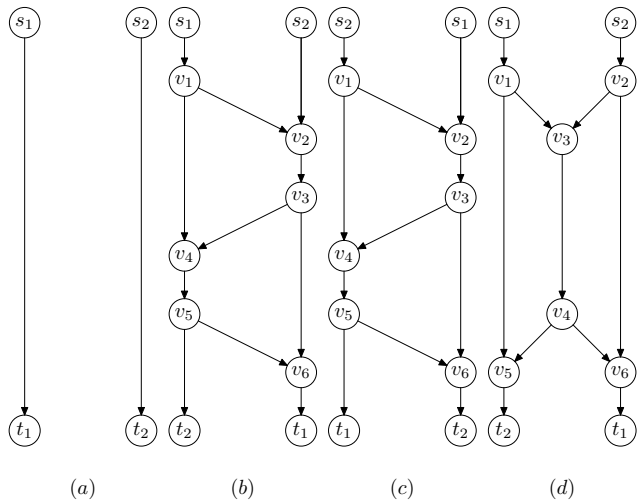

Fig. 1: Four underlying networks of 2-pair unicast networks.

Our method is based on the following two steps: Firstly, decompose a $n$-source $m$-sink network into $nm$ point-to-point subnetworks (for the 2-pair network, $n = m = 2$). Since the properties of a point-to-point network can be easily inferred, this step simplify the initial multi-source multi-sink network coding problem. Secondly, consider the cut set relations of these point-to-point subnetworks. Such relations are shown to contain valuable information of the whole network structure. The first step can simplify the initial problem and the second step can yield a global picture of the original network. A number of "path operations" are used in this paper. That is, a desired path is usually constructed by joining a number of path sections, and conversely, a path will be decomposed into different sections according to particular demands. Our method finally provides an efficient cut-based algorithm to determine the solvability of the 2-pair unicast problem.

The rest of the paper is organized as follows. In Section II, some notations and results which will be used in the sequel are given. The underlying structure of the 2-pair network is presented in Section III. The solvability of the 2-pair unicast problem is analyzed in Section IV. The paper is concluded in Section V.

---

[1] When we finished the first version of this paper, another independent work by Chih-Chun Wang and Ness B. Shroff [18] was published in the ISIT 2007 proceedings. They also derived the four configurations of Fig.1 and presented another characterization as well as a polynomial algorithm of using pebbling games to determine the solvability of 2-pair unicast networks.

## II. Preliminaries

In this paper, most of the discussions are from a graph theoretical point of view. As a preparation, we introduce some basic definitions as well as some simple but frequently used results in this section.

### A. Communication Network, Minimum Cut, and $\mathcal{A}$-Set

A *communication network* $\mathcal{N} = (V, E, S, T)$ consists of a directed acyclic graph (DAG) $G = (V, E)$, a source node set $S \subseteq V$, a sink node set $T \subseteq V$, and a nonnegative capacity $c(e)$ for each link $e \in E$. When $S = \{s\}$ and $T = \{t\}$, i.e., the network has a single source node and a single sink node, it is called *a point-to-pint network* and denoted by $(V, E, s, t)$. Given $s_i \in S$ and $t_j \in T$, it yields a point-to-point network $\mathcal{N}_{i,j} = (V, E, s_i, t_j)$ by considering the other source and sink nodes as internal nodes. Thus there are totally $|S| \times |T|$ point-to-point networks underlying the network $\mathcal{N} = (V, E, S, T)$.

Let $\mathcal{N} = (V, E, s, t)$ be a point-to-point network and let $V = A \cup \bar{A}$ be a vertex partition of $G = (V, E)$ such that $s \in A$ and $t \in \bar{A} = V \setminus A$. An *s-t cut* $C$ is a collection of all the edges from $A$ to $\bar{A}$. The capacity of $C$ is defined as $\sum_{e \in C} c(e)$. The minimum of the cut capacities for all $s$-$t$ cuts is called the *minimum cut capacity* and denoted by $C_\mathcal{N}(s, t)$ or $C(s, t)$ when there is no ambiguity. *A minimum cut* is a cut with the minimum cut capacity. Noticing that there may be a number of minimum cuts within a point-to-point network, the union of those minimum cuts is called the $\mathcal{A}$-*set* (or the *cut set*) of the network (see [20]). Note that the $\mathcal{A}$-set plays an important role in this work.

In this paper, the edges of the network are assumed to have unit capacity, i.e., $c(e) = 1$. In this case, the well-known Max-flow Min-cut Theorem indicates that the *maximum flow* $f$, i.e., the number of edge-disjoint paths from $s$ to $t$ equals to $C(s, t)$, the minimum cut capacity. We call a family of $k$ ($k \in \mathbb{N}$) edge-disjoint paths with common source and sink nodes as an *edge-disjoint k-path*, and denote it by $P^{(k)}$. For a point-to-point network $(V, E, s, t)$ with the maximum flow $f$, it may generally have a number of edge-disjoint $f$-paths from $s$ to $t$. Those edge-disjoint $f$-paths will be denoted by $P_1^{(f)}$, $P_2^{(f)}$, and so on.

*Proposition 2.1:* Let $\mathcal{N} = (V, E, s, t)$ be a point-to-point network with maximum flow $f$. Let $P_1^{(f)}$, $P_2^{(f)}$, $\cdots$, and $P_k^{(f)}$ be all the edge-disjoint $f$-paths from $s$ to $t$. Then we have

$$\mathcal{A} = \bigcap_{i=1}^{k} P_i^{(f)},$$

where $\mathcal{A}$ is the $\mathcal{A}$-set of $\mathcal{N}$ and $P_i^{(f)}$ is considered as the collection of its edges.

*Proof:* Let $e \in \mathcal{A}$. Then there exist a minimum cut $C = \{e_1, e_2, \cdots, e_f\}$ such that $e \in C$. Let $V = A \cup \bar{A}$ be the vertex partition corresponding to $C$. Since $C$ consists of all the edges from $A$ to $\bar{A}$, each path of $P_i^{(f)}$ intersects $C$ for any $i = 1, 2, \cdots, k$. The edge-disjoint condition yields $|P_i^{(f)} \cap C| = f$. Since $|C| = f$, we have $C \subset P_i^{(f)}$ for $i = 1, 2, \cdots, k$, and thus $e \in \bigcap_{i=1}^{k} P_i^{(f)}$. Therefore $\mathcal{A} \subseteq \bigcap_{i=1}^{k} P_i^{(f)}$.

On the other hand, let $e \in \bigcap_{i=1}^{k} P_i^{(f)}$ and consider $\mathcal{N}' = \mathcal{N} \setminus \{e\}$, the network deduced by deleting $e$ from $\mathcal{N}$. We declare that $C_{\mathcal{N}'}(s, t) = f - 1$. In fact, if $C_{\mathcal{N}'}(s, t) = f$, then there will be an edge-disjoint $f$-path from $s$ to $t$ which does not pass through $e$, which contradicts to the assumption $e \in \bigcap_{i=1}^{k} P_i^{(f)}$. Also, $C_{\mathcal{N}'}(s,t)$ can not be less than $f-1$ since $\mathcal{N}'$ is formed by deleting just one edge from $\mathcal{N}$. Now take a minimum cut $C' = \{e'_1, e'_1, \cdots, e'_{f-1}\}$ of $\mathcal{N}'$ and let $V = B \cup \bar{B}$ be the vertex partition corresponding to $C'$. Consider the tail and the head of the edge $e$, denoted by $tail(e)$ and $head(e)$, respectively. If both of them are in $B$, or both are in $\bar{B}$, or $tail(e) \in \bar{B}$ and $head(e) \in B$, then $C'$ also yields a cut of $\mathcal{N}$, which contradicts to that $C_{\mathcal{N}}(s,t) = f$. Thus $tail(e) \in B$ and $head(e) \in \bar{B}$, which implies that $\{e\} \cup C'$ is a (minimum) cut of $\mathcal{N}$. Hence $e \in \mathcal{A}$ which gives $\bigcap_{i=1}^{k} P_i^{(f)} \subseteq \mathcal{A}$. ∎

Obviously, a 2-source 2-sink network yields four point-to-point networks $\mathcal{N}_{i,j} = (V, E, s_i, t_j)$, for $i, j = 1, 2$. In the following part, we use $\mathcal{A}_{i,j}$ to denote the $\mathcal{A}$-set of $\mathcal{N}_{i,j}$.

### B. 2-pair Unicast Network Coding Problem

*Definition 2.2:* A 2-pair unicast problem is specified as follows.
1) A communication network $\mathcal{N} = (V, E, \{s_1, s_2\}, \{t_1, t_2\})$.
2) Two desired unit flows from $s_i$ to $t_i$ for $i = 1, 2$.

Note that the underlying network $\mathcal{N} = (V, E, \{s_1, s_2\}, \{t_1, t_2\})$ is usually called *a 2-pair (unicast) network* in this paper. The desired flows, which are generated in $s_i$ and to be recovered in $t_i$, for $i = 1, 2$, are considered as independent random variables with unit entropies and denoted by $X_1$ and $X_2$, respectively. The information transformation is assumed to be delay-free and error-free. The information transmitted over an edge $e$ and an edge set $A$ are considered as random variables and denoted by $X_e$ and $X_A$, respectively. The entropies of $X_e$ and $X_A$ are simply denoted by $H(e)$ and $H(A)$, respectively.

Without loss of generality, we add an auxiliary source node with a single out-edge (denoted by $S(i)$ for $i = 1, 2$) to each source node and add an auxiliary sink node with a single in-edge (denoted by $T(i)$ for $i = 1, 2$) from each sink node. For convenience, the edges of $S(i)$ and $T(i)$ are called *the information edges*, since they are responsible for delivering and/or recovering the original information. Thus in this paper, each source node $s_i$ is assumed to have one out-edge and no in-edge, and each sink node $t_i$ is assumed to have one in-edge and no out-edge. We also assume that each node except $s_i$ and $t_i$, for $i = 1, 2$, has at least one in-edge and one out-edge.

The information edges $S(i)$ and $T(i)$ can be assumed to have capacity $C(s_i, t_i)$ in order to maintain the maximum flows from $s_i$ to $t_i$ for $i = 1, 2$. But in Section IV-B, information edges are assumed to have unit capacity since the desired information flows have unit rates. Except for the information edges, all the other edges are assumed to have unit capacity.

A *network code* assigned to a 2-pair unicast network $\mathcal{N} = (V, E, \{s_1, s_2\}, \{t_1, t_2\})$ is defined as a collection of functions $\{f_e : e \in E\}$ such that $X_e = f_e(X_{In(e)})$, where $In(e) = \{e' \in E : head(e') = tail(e)\}$ (when $e = S(i)$, then $In(e) = \emptyset$, and let $X_{S(i)} = X_i$ for $i = 1, 2$). A *network coding solution* for a 2-pair unicast network is a network code such that $H(S(i)|T(i)) = 0$ for $i = 1, 2$. A 2-pair unicast problem is called *solvable* when a network coding solution exists (the underlying 2-pair unicast network is called *available*), and called *unsolvable* (the underlying 2-pair unicast network is called *unavailable*) otherwise.

*Remark 2.3:* By the definition, for any network code $\{f_e : e \in E\}$, the condition that $X_e$ is a function of $X_{In(e)}$ indicates that $H(X_e|X_{In(e)}) = 0$.

Unlike the definition of a network coding solution in [8], where an arbitrary positive network coding rate is considered, the 2-pair unicast problem here aims at supporting two unit rate flows. Hence, the definition of a network coding solution has been slightly changed. In fact, it corresponds to the network coding solution in [8] with rate $\geq 1$ .

*C. Path Combination/Decomposition*

A (simple) path can be represented as a string of ordered edges, $P = (e_1, e_2, \cdots, e_n)$, with $head(e_i) = tail(e_{i+1})$ for $i = 1, 2, \cdots, n-1$, where $e_i$ is called an *up-link* (*down-link*) of $e_j$ if $i < j$ ($i > j$). We use $e \in P$ to denote an edge $e$ lies in a path $P$. For a DAG, it is widely known that there exists a *topological order* for the edges according to the up- (or down-) link relation, that is, if $e_i$ is an up-link of $e_j$ for some path $P$, then $e_i$ is an up-link of $e_j$ for any path $Q$ for $e_i, e_j \in Q$. This topological order of the edges of a DAG will always be used in this paper.

A frequently used technique in this paper is path combination/decomposition. We denote $P[v_i, v_j]$ as the section of $P$ from node $v_i$ to node $v_j$. Similarly, $P[e_i, e_j]$ is used to denote the section of $P$ from $tail(e_i)$ to $head(e_j)$, where $e_i$ and $e_j$ are two different edges in $P$. We also use $P[e_i, v_j]$ and $P[v_i, e_j]$ to denote the sections of $P$ from $tail(e_i)$ to node $v_j$, and from node $v_i$ to $head(e_j)$, respectively. Let $P_1 = (e_1, e_2, \cdots, e_n)$ and $P_2 = (e'_1, e'_2, \cdots, e'_m)$ be two paths such that $head(P_1) = tail(P_2)$ (that is, $head(e_n) = tail(e'_1)$). We denote the path $P = (e_1, e_2, \cdots, e_n, e'_1, \cdots, e'_m)$ as $P_1$-$P_2$. Similarly, we use $P$-$P^{(k)}$ to denote the configuration by joining a simple path $P$ and an edge-disjoint $k$-path $P^{(k)}$. An edge-disjoint $k$-path composed by $s$-$t$ paths $P_1, P_2, \cdots, P_k$ is sometimes denoted as $P^{(k)} = P_1 \cup P_2 \cup \cdots \cup P_k$. Moreover, a path is usually regarded as a collection of edges. For example, we use $P \cup Q$ and $P \cap Q$ to represent the union and the intersection ( of the edges ) of paths $P$ and $Q$, respectively.

### III. NETWORK STRUCTURE ANALYSIS

In this section, we explore the underlying structure of 2-pair unicast networks. In the following, the 2-pair network will be assumed with $C(s_1, t_1) = C(s_2, t_2) = 1$. For the case $C(s_1, t_1) \cdot C(s_2, t_2) \geq 2$, it will be discussed later (If $C(s_1, t_1) \cdot C(s_2, t_2) = 0$, then there is no path from $s_1$ to $t_1$ or from $s_2$ to $t_2$, and the 2-pair unicast problem is unsolvable obviously.).

Throughout the paper, the terms "$\mathcal{N}$ has underlying network $\mathcal{N}_0$," " $\mathcal{N}$ contains a copy of $\mathcal{N}_0$," or simply "$\mathcal{N}$ contains $\mathcal{N}_0$" will be equivalently used to indicate the existence of a same topology between paths of $\mathcal{N}$ and edges of $\mathcal{N}_0$. Formally, we give the following definition.

*Definition 3.1:* Let $\mathcal{N} = (V, E, \{s_1, t_1\}, \{s_2, t_2\})$ and $\mathcal{N}_0 = (V', E', \{s'_1, t'_1\}, \{s'_2, t'_2\})$ be two 2-pair unicast networks. We say $\mathcal{N}$ contains a copy of $\mathcal{N}_0$ if there exists a function $f$ from the edges of $\mathcal{N}_0$ to the paths of $\mathcal{N}$ satisfying:

(1) If $tail(e') = s'_i$, then $tail(f(e')) = s_i$, for $e' \in E'$ and $i = 1, 2$;

(2) If $head(e') = t'_i$, then $head(f(e')) = t_i$, for $e' \in E'$ and $i = 1, 2$;

(3) If $head(e'_1) = tail(e'_2)$, then $head(f(e'_1)) = tail(f(e'_2))$, for $e'_1, e'_2 \in E'$;

(4) If $e'_1 \neq e'_2$, then $f(e'_1)$ and $f(e'_2)$ are edge-disjoint, for $e'_1, e'_2 \in E'$.

Obviously, this definition can be generalized to an arbitrary multi-source multi-sink network, as similar to the notion of *subgraph homeomorphism* in graph theory (see [14]). Generally, paths under the subgraph homeomorphism are needed to be node-disjoint, which is naturally loosened here to edge-disjoint since the network information flow problem concentrates on the link capacity constrains. Before illustrating the main results, we give a lemma.

*Lemma 3.2:* Let $\mathcal{N} = (V, E, s, t)$ be a point-to-point network such that $C(s,t) = 1$. Denote $\mathcal{A}$ as its $\mathcal{A}$-set. Assume that $s$ has a unique out-edge, $S(1)$, and $t$ has a unique in-edge, $T(1)$. Then the following items hold.

1) For any edge $e \in \mathcal{A}$ and any $s$-$t$ path $P$, we have $e \in P$;
2) For edge $e \notin \mathcal{A}$, there exists an $s$-$t$ path $P$ such that $e \notin P$;
3) $\mathcal{N}$ has a subnetwork $\mathcal{N}_0 = P_1\text{-}P_1^{(2)}\text{-}P_2\text{-}P_2^{(2)}\text{-}\cdots\text{-}P_n^{(2)}\text{-}P_{n+1}$ such that $\mathcal{A} = P_1 \cup P_2 \cup \cdots \cup P_{n+1}$, where $tail(P_1) = s$, $head(P_{n+1}) = t$, and path $P_i$ is regarded as the collection of edges.

*Proof:* The first two items are obvious by Proposition 2.1. Now we prove 3) by constructing $\mathcal{N}_0$. Let $\mathcal{A} = \{e_1, e_2, \cdots, e_m\}$ such that $e_i$ is an up-link of $e_j$ for $i < j$ (where, $e_1 = S(1)$, and $e_m = T(1)$). Let $e_i, e_{i+1} \in \mathcal{A}$ and $head(e_i) \neq tail(e_{i+1})$. Note that $(V, E, head(e_i), tail(e_{i+1}))$ is also a point-to-point network.

Consider $C(head(e_i), tail(e_{i+1}))$. If $C(head(e_i), tail(e_{i+1})) = 1$, then there exists a $head(e_i)$-$tail(e_{i+1})$ minimum cut which contains only one edge, namely, $\{e\}$. Since $e$ is a down-link of $e_i$ and an up-link of $e_{i+1}$, we have $e \notin \mathcal{A}$. On the other hand, by Proposition 2.1, any $head(e_i)$-$tail(e_{i+1})$ path must pass through $e$. Thus any $s$-$t$ path must pass through $e$. By Proposition 2.1, $e \in \mathcal{A}$, which contradicts to $e \notin \mathcal{A}$. Therefore $C(head(e_i), tail(e_{i+1})) \geq 2$.

Take an edge-disjoint 2-path from $head(e_i)$ to $tail(e_{i+1})$, and denote it as $Q_i^{(2)}$. Suppose that $e_{i_1}, e_{i_2}, \cdots, e_{i_n}$ are all the links of $\mathcal{A}$ with $head(e_{i_k}) \neq tail(e_{i_k+1})$. Let $P_1 = (e_1, \cdots, e_{i_1})$, $P_k = (e_{i_{k-1}+1}, \cdots, e_{i_k})$, for $k = 2, 3, \cdots, n$, $P_{n+1} = (e_{i_n+1}, \cdots, e_m)$, and $P_k^{(2)} = Q_{i_k}^{(2)}$ for $k = 1, 2, \cdots, n$. Then $\mathcal{N}_0 = P_1\text{-}P_1^{(2)}\text{-}P_2\text{-}P_2^{(2)}\text{-}\cdots\text{-}P_n^{(2)}\text{-}P_{n+1}$ satisfies the desired conditions. The proof is done. ∎

By observing the proof process of Lemma 3.2, we get the following corollary.

*Corollary 3.3:* Let $\mathcal{N} = (V, E, s, t)$ be a point-to-point network such that $C(s,t) = 1$ with $\mathcal{A}$-set $\mathcal{A} = \{e_1, e_2, \cdots, e_n\}$. If $head(e_i) \neq tail(e_{i+1})$ for some $1 \leq i < n$, then there exists an edge-disjoint 2-path from $head(e_i)$ to $tail(e_{i+1})$.

Now we start to discuss the characteristics of a 2-pair unicast network with $\mathcal{A}_{1,1} \cap \mathcal{A}_{2,2} = \emptyset$. Note that $\mathcal{A}_{i,j}$ is the $\mathcal{A}$-set of the point-to-point network $\mathcal{N}_{i,j} = (V, E, s_i, t_j)$, for $i, j = 1, 2$.

*Theorem 3.4:* Let $\mathcal{N} = (V, E, \{s_1, t_1\}, \{s_2, t_2\})$ be a 2-pair unicast network with $\mathcal{A}_{1,1} \cap \mathcal{A}_{2,2} = \emptyset$. Then there is either an $s_1$-$t_1$ path disjoint with $\mathcal{A}_{2,2}$ or an $s_2$-$t_2$ path disjoint with $\mathcal{A}_{1,1}$.

*Proof:* Let $\mathcal{A}_{1,1} = \{e_1, e_2, \cdots, e_n\}$ such that $e_i$ is an up-link of $e_j$ for $i < j$ and let $P_1$ be an $s_1$-$t_1$ path. If $P_1 \cap \mathcal{A}_{2,2} = \emptyset$, then we are done. Now suppose $P_1$ contains an edge $e^* \in \mathcal{A}_{2,2}$. Fix $m$, $0 \leq m \leq n$, such that $e^*$ is a down-link of $e_i$ for $i \leq m$ and an up-link of $e_i$ for $i > m$. We can construct an $s_2$-$t_2$ path disjoint with $\mathcal{A}_{1,1}$ as follows.

If $m > 0$, then we can find an $s_2$-$t_2$ path $P_2$ not containing $e_m$ since $e_m \notin \mathcal{A}_{2,2}$. ( if $m = 0$, then $P_2$ can be any $s_2$-$t_2$ path.) Since $e^* \in \mathcal{A}_{2,2}$, $e^*$ lies on $P_2$. The path $P_2[s_2, e^*]$ cannot contain edges $e_i$ for $i < m$, because, if it did, then $P_1[s_1, e_i]$-$P_2[head(e_i), e^*]$-$P_1[head(e^*), t_1]$ would be an $s_1$-$t_1$ path not containing $e_m$, which contradicts to $e_m \in \mathcal{A}_{1,1}$. Also $P_2[s_2, e^*]$ cannot contain any edge $e_j$ with $j > m$ because this would make $e_j$ an up-link of $e^*$ in $P_2$ and a down-link of $e^*$ in $P_1$. Thus $P_2[s_2, e^*] \cap \mathcal{A}_{1,1} = \emptyset$.

Similarly, if $m < n$, we can find an $s_2$-$t_2$ path $P_2'$ not containing $e_{m+1}$. (If $m = n$, $P_2'$ can be any $s_2$-$t_2$ path.) A similar argument as above shows that $P_2'[e^*, t_2] \cap \mathcal{A}_{1,1} = \emptyset$.

Combining $P_2[s_2, e^*]$ and $P_2'[e^*, t_2]$ together, we have an $s_2$-$t_2$ path $P_2[s_2, e^*]$-$P_2'[head(e^*), t_2]$, which is disjoint with $\mathcal{A}_{1,1}$. The proof is completed. ∎

*Theorem 3.5:* Let $\mathcal{N} = (V, E, \{s_1, t_1\}, \{s_2, t_2\})$ be a 2-pair unicast network. If $\mathcal{A}_{1,1} \cap \mathcal{A}_{2,2} = \emptyset$, then the network contains Fig.1(a), Fig.1(b), or Fig.1(c).

*Proof:* By Theorem 3.4, we first assume that there exists an $s_2$-$t_2$ path disjoint with $\mathcal{A}_{1,1}$, and prove that the network contains Fig.1(a) or Fig.1(b).

By Lemma 3.2, let $\mathcal{N}_0 = P_1$-$P_1^{(2)}$-$P_2$-$P_2^{(2)}$-$\cdots$-$P_{n-1}^{(2)}$-$P_n$ be a subnetwork of $\mathcal{N}$ such that $\mathcal{A}_{1,1} = P_1 \cup P_2 \cup \cdots \cup P_n$ with $P_i^{(2)} = Q_i \cup Q_i'$ for $i = 1, 2, \cdots, n-1$. Let $P$ be an $s_2$-$t_2$ path disjoint with $\mathcal{A}_{1,1}$. If $P \cap \mathcal{N}_0 = \emptyset$, then $\mathcal{N}$ contains Fig.1(a) since $P_1$-$Q_1$-$P_2$-$Q_2$-$\cdots$-$Q_{n-1}$-$P_n$ and $P$ are edge-disjoint $s_1$-$t_1$ and $s_2$-$t_2$ paths. If $P \cap \mathcal{N}_0 \neq \emptyset$, then assume $e^* \in P \cap \mathcal{N}_0$ and let $e^* \in Q_m$ for some $1 \leq m \leq n-1$. We now prove that $\mathcal{N}$ contains Fig.1(a) or Fig.1(b).

We claim first that $P \cap P_i^{(2)} = \emptyset$ for $i \neq m$. If it is not true, without loss of generality, assume $e' \in P \cap Q_i$ and consider the following two cases. (1) $i < m$. Since $e'$ is an up-link of $e^*$ in $P_1$-$Q_1$-$P_2$-$Q_2$-$\cdots$-$Q_{n-1}$-$P_n$, $e'$ is an up-link of $e^*$ according to $P$. So $P_1$-$Q_1$-$\cdots$-$Q_i[tail(Q_i), e']$-$P[e', e^*]$-$Q_m[e^*, head(Q_m)]$-$P_{m+1}$-$\cdots$-$P_n$ is an $s_1$-$t_1$ path disjoint with $P_m$, which contradicts to $P_m \subset \mathcal{A}_{1,1}$. (2) $i > m$. Similarly, one can see that $s_1$-$t_1$ path $P_1$-$Q_1$-$\cdots$-$Q_m[tail(Q_m), e^*]$-$P[e^*, e']$-$Q_i[e', head(Q_i)]$-$P_{i+1}$-$\cdots$-$P_n$ is disjoint with $P_i \subset \mathcal{A}_{1,1}$, a contradiction.

Now assume that $\mathcal{N}_0 \cap P = P_m^{(2)} \cap P = (Q_m \cup Q_m') \cap P = \{e_1, e_2, \cdots, e_r\}$ such that $e_j$ is a down-link of $e_i$ for $1 \leq i < j \leq r$. Consider the following cases:

1) If $e_1, e_r \in Q_m$, as shown in Fig.2(a), then $s_2$-$t_2$ path $P[s_2, tail(e_1)]$-$Q_m[e_1, e_r]$-$P[head(e_r), t_2]$ is edge-disjoint with $s_1$-$t_1$ path $P_1$-$Q_1$-$\cdots$-$P_m$-$Q_m'$-$P_{m+1}$-$\cdots$-$Q_{n-1}$-$P_n$. The network contains Fig.1(a).

2) If $e_1 \in Q_m$ and $e_r \in Q_m'$, then let $k$ be an index such that $e_k \in Q_m$, and $e_{k'} \in Q_m'$ for all $k' > k$, as shown in Fig.2(b). It can be checked that the network contains Fig.1(b) with the function $f$: $(s_1, v_1) \mapsto P_1$-$Q_1$-$\cdots$-$P_m$; $(s_2, v_2) \mapsto P[s_2, tail(e_1)]$; $(v_6, t_1) \mapsto P_{m+1}$-$Q_{m+1}$-$\cdots$-$P_n$; $(v_5, t_2) \mapsto P[head(e_r), t_2]$; $(v_1, v_2) \mapsto Q_m[tail(Q_m), tail(e_1)]$; $(v_1, v_4) \mapsto Q_m'[tail(Q_m'), tail(e_{k+1})]$; $(v_2, v_3) \mapsto Q_m[tail(e_1), head(e_k)]$; $(v_3, v_4) \mapsto P[head(e_k), tail(e_{k+1})]$; $(v_4, v_5) \mapsto Q_m'[tail(e_{k+1}), head(e_r)]$; $(v_5, v_6) \mapsto Q_m'[head(e_r), head(Q_m')]$. The imaged paths are edge-disjoint because any two disjoint sections of $P$ ( and $\mathcal{N}_0$ ) are edge-disjoint, and $P \cap \mathcal{N}_0 = P \cap P_m^{(2)}$.

3) If $e_1, e_r \in Q_m'$, the discussion is similarly to that of case 1). The network contains Fig.1(a).

4) If $e_1 \in Q_m'$ and $e_r \in Q_m$, the discussion is similar to that of case 2). The network contains Fig.1(b).

In the case where there exists an $s_1$-$t_1$ path disjoint with $\mathcal{A}_{2,2}$, one can prove symmetrically that the network contains Fig.1(a) or Fig.1(c). ∎

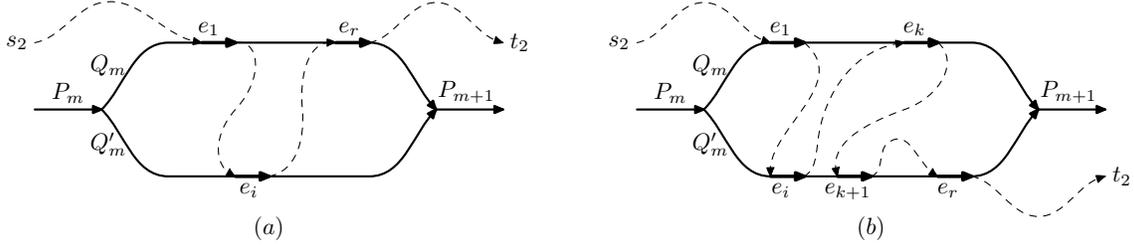

Fig. 2: The relationship between $P$ and $\mathcal{N}_0$.

(a): The case of $e_1, e_r \in Q_m$. (b): The case of $e_1 \in Q_m, e_r \in Q'_m$.

In the above discussions, we have deduced the underlying structure of the 2-pair unicast network with $\mathcal{A}_{1,1} \cap \mathcal{A}_{2,2} = \emptyset$. Now we deal with the 2-pair networks with $\mathcal{A}_{1,1} \cap \mathcal{A}_{2,2} \neq \emptyset$. Firstly, we need a lemma.

*Lemma 3.6:* Let $\mathcal{N} = (V, E, \{s_1, t_1\}, \{s_2, t_2\})$ be a 2-pair unicast network such that $\mathcal{A}_{1,1} \cap \mathcal{A}_{2,2} \neq \emptyset$, and let $\mathcal{A}_{1,1} = \{e_1, e_2, \cdots, e_n\}$. If $e_i, e_j \in \mathcal{A}_{1,1} \cap \mathcal{A}_{2,2}$ ($i < j$), then $e_\ell \in \mathcal{A}_{1,1} \cap \mathcal{A}_{2,2}$ for $i < \ell < j$.

*Proof:* Assume $e_\ell \notin \mathcal{A}_{2,2}$ and let $Q$ be an $s_2$-$t_2$ path not containing $e_\ell$. Then, for any $s_1$-$t_1$ path $P$, we have an $s_1$-$t_1$ path $P' = P[s_1, tail(e_i)]$-$Q[e_i, e_j]$-$P[head(e_j), t_1]$ not containing $e_\ell$. Therefore $e_\ell \notin \mathcal{A}_{1,1}$, a contradiction. ∎

Given a 2-pair unicast network $\mathcal{N} = (V, E, \{s_1, t_1\}, \{s_2, t_2\})$, an $s_1$-$t_1$ path $P$ and an $s_2$-$t_2$ path $Q$, by Lemma 3.2, one can have that $P \supseteq \mathcal{A}_{1,1}$ and $Q \supseteq \mathcal{A}_{2,2}$, and thus $P \cap Q \supseteq \mathcal{A}_{1,1} \cap \mathcal{A}_{2,2}$. Moreover, when $\mathcal{A}_{1,1} \cap \mathcal{A}_{2,2} \neq \emptyset$, one can prove further that there exist an $s_1$-$t_1$ path $P$ and an $s_2$-$t_2$ path $Q$ such that $P \cap Q = \mathcal{A}_{1,1} \cap \mathcal{A}_{2,2}$ as the following theorem shows.

*Theorem 3.7:* Let $\mathcal{N} = (V, E, \{s_1, t_1\}, \{s_2, t_2\})$ be a 2-pair unicast network such that $\mathcal{A}_{1,1} \cap \mathcal{A}_{2,2} \neq \emptyset$. Then there exist an $s_1$-$t_1$ path $P$ and an $s_2$-$t_2$ path $Q$ such that $P \cap Q = \mathcal{A}_{1,1} \cap \mathcal{A}_{2,2}$.

*Proof:* We construct $P$ and $Q$ by using the technique of path combination (see Fig.3(a)). By Lemma 3.6, one can let $\mathcal{A}_{1,1} = \{e_1, e_2, \cdots, e_n\}$ and $\mathcal{A}_{1,1} \cap \mathcal{A}_{2,2} = \{e_m, e_{m+1}, \cdots, e_{m+j}\}$. Moreover, we have that $m \geq 2$ and $m + j \leq n - 1$ by the assumptions that $s_i$ has a single out-edge and $t_i$ has a single in-edge. Denote $tail(e_m) = s$ and $head(e_{m+j}) = t$. We claim that there exist an $s_1$-$s$ path $\hat{P}$ and an $s_2$-$s$ path $\hat{Q}$ such that $\hat{P} \cap \hat{Q} = \emptyset$.

To prove this, let $\hat{P}'$ be an arbitrary $s_1$-$t_1$ path. By Lemma 3.2, one can take an $s_2$-$t_2$ path $\hat{Q}'$ such that $e_{m-1} \notin \hat{Q}'$. If there is an $e^* \in \hat{P}'[s_1, e_{m-1}] \cap \hat{Q}'[s_2, s]$, then $\hat{P}'[s_1, e^*]$-$\hat{Q}'[head(e^*), s]$-$\hat{P}'[s, t_1]$ is an $s_1$-$t_1$ path not containing $e_{m-1}$. So $e_{m-1} \notin \mathcal{A}_{1,1}$, resulting in a contradiction. Thus there exist an $s_1$-$head(e_{m-1})$ path $\hat{P}'[s_1, e_{m-1}]$ and an $s_2$-$s$ path $\hat{Q}'[s_2, s]$ with $\hat{P}'[s_1, e_{m-1}] \cap \hat{Q}'[s_2, s] = \emptyset$. If $head(e_{m-1}) = s$, then we are done by letting $\hat{P} = \hat{P}'[s_1, s]$ and $\hat{Q} = \hat{Q}'[s_2, s]$. Now suppose that $head(e_{m-1}) \neq s$. By Lemma 3.2, there exists an edge-disjoint 2-path $P^{(2)}$ from $head(e_{m-1})$ to $s$. Let $P^{(2)} = Q_1 \cup Q_2$. If $\hat{Q}' \cap P^{(2)} = \emptyset$, then we are done by letting $\hat{P} = \hat{P}'[s_1, e_{m-1}]$-$Q_1$ and $\hat{Q} = \hat{Q}'[s_2, s]$. If $\hat{Q}' \cap P^{(2)} \neq \emptyset$, let $\{\tilde{e}_1, \tilde{e}_2, \cdots, \tilde{e}_\ell\} \in \hat{Q}' \cap P^{(2)}$ such that $\tilde{e}_j$ is a down-link of $\tilde{e}_i$ for $i < j$. Without loss of generality, assume that $\tilde{e}_1 \in Q_1$. Then $\hat{P} = \hat{P}'[s_1, e_{m-1}]$-$Q_2$ is an $s_1$-$s$ path and $\hat{Q} = \hat{Q}'[s_2, \tilde{e}_1]$-$Q_1[head(\tilde{e}_1), s]$ is an $s_2$-$s$ path satisfy $\hat{P} \cap \hat{Q} = \emptyset$.

Similarly, one can find a $t$-$t_1$ path $\check{P}$ and a $t$-$t_2$ path $\check{Q}$ with $\check{P} \cap \check{Q} = \emptyset$.

Let $e_{i_1}, e_{i_2}, \cdots, e_{i_n}$ be all the links such that $head(e_{i_k}) \neq tail(e_{i_k+1})$ for $m \leq i_k < m + j$. Noticing that $e_{i_k}, e_{i_k+1} \in \mathcal{A}_{1,1}$, there exist an edge-disjoint 2-path, namely, $\bar{P}_k^{(2)} = \bar{Q}_k \cup \bar{Q}'_k$ from $head(e_{i_k})$ to $tail(e_{i_k+1})$ by Corollary 3.3. Let $\bar{P}_1 = (e_m, \cdots, e_{i_1})$, $\bar{P}_k = (e_{i_{k-1}+1}, \cdots, e_{i_k})$ for $k = 2, 3, \cdots, n$, and $\bar{P}_{n+1} = (e_{i_n+1}, \cdots, e_{m+j})$. Set $\bar{P}=\bar{P}_1\text{-}\bar{Q}_1\text{-}\bar{P}_2\text{-}\bar{Q}_2\text{-}\cdots\text{-}\bar{Q}_n\text{-}\bar{P}_{n+1}$ and $\bar{Q}=\bar{P}_1\text{-}\bar{Q}'_1\text{-}\bar{P}_2\text{-}\bar{Q}'_2\text{-}\cdots\text{-}\bar{Q}'_n\text{-}\bar{P}_{n+1}$. We have $\bar{P} \cap \bar{Q} = \{e_m, e_{m+1}, \cdots, e_{m+j}\} = \mathcal{A}_{1,1} \cap \mathcal{A}_{2,2}$.

Let $P = \hat{P}\text{-}\bar{P}\text{-}\check{P}$ and $Q = \hat{Q}\text{-}\bar{Q}\text{-}\check{Q}$. Then $P$ is an $s_1$-$t_1$ path and $Q$ is an $s_2$-$t_2$ path such that $P \cap Q = \mathcal{A}_{1,1} \cap \mathcal{A}_{2,2}$, which completes the proof. ■

*Corollary 3.8:* Let $\mathcal{N} = (V, E, \{s_1, t_1\}, \{s_2, t_2\})$ be a 2-pair unicast network such that $\mathcal{A}_{1,1} \cap \mathcal{A}_{2,2} \neq \emptyset$. Then $\mathcal{N}$ contains a copy of the network as shown in Fig.3(b).

*Proof:* Using the notations in the proof of Theorem 3.7, a function $f$ can be assigned from the edges of Fig.3(b) to the paths of Fig.3(a) such that $(s_1, v_1) \mapsto \hat{P}$; $(s_2, v_1) \mapsto \hat{Q}$; $(v_1, v_2) \mapsto \bar{P}$; $(v_2, t_1) \mapsto \check{P}$; and $(v_2, t_2) \mapsto \check{Q}$. The imaged paths are edge-disjoint because: (1) the edges in $\hat{P}$ and in $\hat{Q}$ are up-links of the edges in $\bar{P}$; (2) the edges in $\bar{P}$ are up-links of the edges in $\check{P}$ and in $\check{Q}$; and (3) $\hat{P} \cap \hat{Q} = \check{P} \cap \check{Q} = \emptyset$. ■

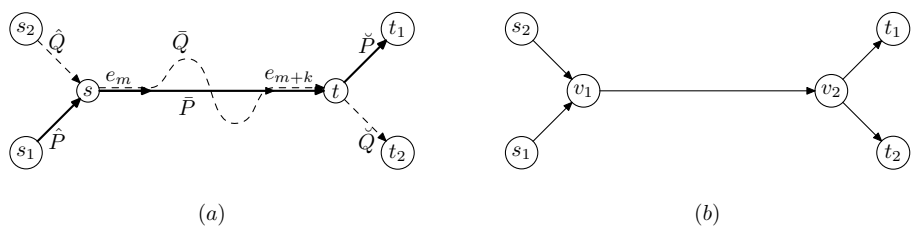

Fig. 3: The case $\mathcal{A}_{1,1} \cap A_{2,2} \neq \emptyset$.

($a$): The construction of paths $P$ and $Q$ such that $P \cap Q = \mathcal{A}_{1,1} \cap \mathcal{A}_{2,2}$ with $P$ in bold line and $Q$ in dashed line.

($b$) : The underlying network for the 2-pair unicast network with $\mathcal{A}_{1,1} \cap \mathcal{A}_{2,2} \neq \emptyset$.

Based on Theorem 3.7, we have the following theorem.

*Theorem 3.9:* Let $\mathcal{N} = (V, E, \{s_1, t_1\}, \{s_2, t_2\})$ be a 2-pair unicast network such that $\mathcal{A}_{1,1} \cap \mathcal{A}_{2,2} \neq \emptyset$ and there exist an $s_1$-$t_2$ path $P_1$ and an $s_2$-$t_1$ path $P_2$ with $P_i \cap (\mathcal{A}_{1,1} \cap \mathcal{A}_{2,2}) = \emptyset$ for $i = 1, 2$. Then $\mathcal{N}$ contains Fig.1(d).

*Proof:* Let $\mathcal{A}_{1,1} \cap \mathcal{A}_{2,2} = \{e_1, e_2, \cdots, e_k\}$. By Theorem 3.7, there exist an $s_1$-$t_1$ path $P$ and an $s_2$-$t_2$ path $Q$ such that $P \cap Q = \{e_1, e_2, \cdots, e_k\}$. Let $P_1$ be an $s_1$-$t_2$ path and $P_2$ be an $s_2$-$t_1$ path such that $P_i \cap \mathcal{A}_{1,1} \cap \mathcal{A}_{2,2} = \emptyset$ ( $i = 1, 2$ ). We prove firstly the following properties of $P$, $Q$, $P_1$ and $P_2$ and then prove $\mathcal{N}$ contains Fig.1(d).

1) $P_1 \cap P[e_1, t_1] = \emptyset$, $P_1 \cap Q[s_2, e_k] = \emptyset$;
2) $P_2 \cap P[s_1, e_k] = \emptyset$, $P_2 \cap Q[e_1, t_2] = \emptyset$;
3) $P_1 \cap P_2 = \emptyset$.

We prove them one by one.

1) Suppose that $P_1 \cap P[e_1, t_1] \neq \emptyset$. Let $e \in P_1 \cap P[e_1, t_1]$. Then $P_1[s_1, e]$-$P[head(e), t_1]$ is an $s_1$-$t_1$ path not containing $e_1 \in \mathcal{A}_{1,1}$, resulting in a contradiction. Thus $P_1 \cap P[e_1, t_1] = \emptyset$. Similarly, if $e \in P_1 \cap Q[s_2, e_k]$ for some edge $e$, then $Q[s_2, e]$-$P_1[head(e), t_2]$ is an $s_2$-$t_2$ path without passing through $e_1$, which contradicts to $e_1 \in \mathcal{A}_{2,2}$.

2) It can be proved similarly.

3) Suppose that $P_1 \cap P_2 \neq \emptyset$, and let $e' \in P_1 \cap P_2$. Then, $P_1[s_1, e']$-$P_2[head(e'), t_1]$ is an $s_1$-$t_1$ path not containing $e_1$, which is a contradiction to $e_1 \in \mathcal{A}_{1,1}$.

By property 1), we can assume that $P_1 \cap P = P_1 \cap P[s_1, tail(e_1)] = \{\hat{e}_1, \hat{e}_2, \cdots, \hat{e}_n\}$ and $P_1 \cap Q = P_1 \cap Q[head(e_k), t_2] = \{\breve{e}_1, \breve{e}_2, \cdots, \breve{e}_m\}$, (both in the topological order). It can be seen that (1) $\{\hat{e}_1, \hat{e}_2, \cdots, \hat{e}_n\} \neq \emptyset$ and $\{\breve{e}_1, \breve{e}_2, \cdots, \breve{e}_m\} \neq \emptyset$; and (2) $head(\hat{e}_n) \neq tail(e_1)$ and $head(e_k) \neq tail(\breve{e}_1)$. In fact, (1) holds since $\hat{e}_1 = S(1)$ is the unique out-edge of $s_1$ and $\breve{e}_m = T(2)$ is the unique in-edge of $t_2$. For the property (2), if $head(\hat{e}_n) = tail(e_1)$, then $Q[s_2, tail(e_1)]$-$P_1[head(\hat{e}_n), t_2]$ is an $s_2$-$t_2$ path disjoint with $\mathcal{A}_{2,2}$, while if $head(e_k) = tail(\breve{e}_1)$, then $P_1[s_1, tail(\breve{e}_1)]$-$P[head(e_k), t_1]$ is an $s_1$-$t_1$ path disjoint with $\mathcal{A}_{1,1}$. Both are contradictions.

Similarly, one can prove that $P_2 \cap Q = P_2 \cap Q[s_2, tail(e_1)] \neq \emptyset$ and $P_2 \cap P = P_2 \cap P[head(e_k), t_1] \neq \emptyset$. Let $P_2 \cap Q = P_2 \cap Q[s_2, tail(e_1)] = \{\hat{e}'_1, \hat{e}'_2, \cdots, \hat{e}'_u\}$ and let $P_2 \cap P = P_2 \cap P[head(e_k), t_1] = \{\breve{e}'_1, \breve{e}'_2, \cdots, \breve{e}'_v\}$. We have $head(\hat{e}'_u) \neq tail(e_1)$ and $head(e_k) \neq tail(\breve{e}'_1)$.

Now we can define a function $f$ from the edges of Fig.1(d) to the paths $P, Q, P_1, Q_1$ of $\mathcal{N}$ (see Fig.4): $(s_1, v_1) \mapsto P[s_1, \hat{e}_n]$; $(s_2, v_2) \mapsto Q[s_2, \hat{e}'_u]$; $(v_1, v_3) \mapsto P[head(\hat{e}_n), tail(e_1)]$; $(v_2, v_3) \mapsto Q[head(\hat{e}'_u), tail(e_1)]$; $(v_3, v_4) \mapsto P[e_1, e_k]$; $(v_1, v_5) \mapsto P_1[head(\hat{e}_n), tail(\breve{e}_1)]$; $(v_2, v_6) \mapsto P_2[head(\hat{e}'_u), tail(\breve{e}'_1)]$; $(v_4, v_5) \mapsto Q[head(e_k), tail(\breve{e}_1)]$; $(v_4, v_6) \mapsto P[head(e_k), tail(\breve{e}'_1)]$; $(v_6, t_1) \mapsto P[tail(\breve{e}'_1), t_1]$; $(v_5, t_2) \mapsto Q[tail(\breve{e}_1), t_2]$. Obviously, $f$ results in disjoint paths. The theorem is proved. ∎

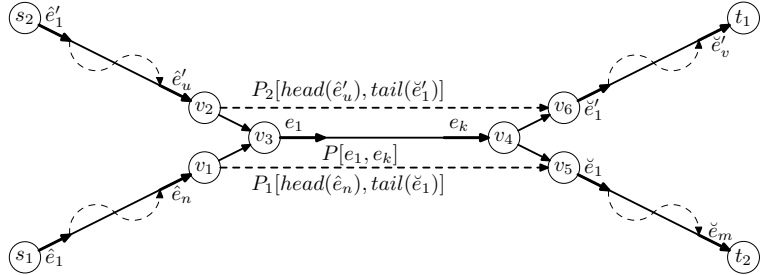

Fig. 4: The figure illustrating the proof of Theorem 3.9.
In the figure, the path sections of $P$ and $Q$ ($P_1$ and $P_2$) are shown in bold (dashed) lines.

We discussed the structures of 2-pair unicast networks with $C(s_1, t_1) \cdot C(s_2, t_2) = 1$ previously. For the network with $C(s_1, t_1) \cdot C(s_2, t_2) \geq 2$, its structure can be deduced directly from Theorem 3.5.

*Corollary 3.10:* Let $\mathcal{N} = (V, E, \{s_1, t_1\}, \{s_2, t_2\})$ be a 2-pair unicast network. If $C(s_1, t_1) \cdot C(s_2, t_2) \geq 2$, then $\mathcal{N}$ contains a copy of the networks Fig.1(a), Fig.1(b), or Fig.1(c).

*Proof:* Without loss of generality, we assume that $C(s_1, t_1) \geq 2$. By the prior assumptions, $s_i$ ($t_i$) has the unique out-edge $S(i)$ (in-edge $T(i)$) with capacity $C(s_i, t_i)$ for $i = 1, 2$, and except for these four edges, all the other edges have unit capacities. Then the Max-flow Min-cut theorem implies that there exist an edge-disjoint 2-path $P^{(2)}$ from $head(S(1))$ to $tail(T(1))$. Let $P^{(2)} = Q \cup Q'$ and take an $s_2$-$t_2$ path $P$. If $P^{(2)} \cap P = \emptyset$, then $\mathcal{N}$ contain Fig.1(a) by noticing that $S(1)$-$Q$-$T(1)$ and $P$ are edge-disjoint. Now assume $P^{(2)} \cap P = \{e_1, e_2, \cdots, e_r\}$. Similar to the latter part of the proof of Theorem 3.5, there are 4 cases need to be discussed (A figure to illustrate these cases is a minor modification on Fig.2 by replacing $Q_m$, $Q'_m$, $P_m$ and $P_{m+1}$ with $Q$, $Q'$, $S(1)$, and $T(1)$ respectively):

1) If $e_1, e_r \in Q$, then $S(1)$-$Q'$-$T(1)$ is an $s_1$-$t_1$ path which is edge-disjoint with the $s_2$-$t_2$ path $P[s_2, tail(e_1)]$-$Q[e_1, e_r]$-$P[head(e_r), t_2]$. The network contains Fig.1(a).

2) If $e_1 \in Q$ and $e_r \in Q'$, let $k$ be the maximum index such that $e_k \in Q$ and $e_{k+1} \in Q'$ and let $f$ be defined as $(s_1, v_1) \mapsto S(1)$; $(s_2, v_2) \mapsto P[s_2, tail(e_1)]$; $(v_6, t_1) \mapsto T(1)$; $(v_5, t_2) \mapsto P[head(e_r), t_2]$; $(v_1, v_2) \mapsto Q[tail(Q), tail(e_1)]$; $(v_1, v_4) \mapsto Q'[tail(Q'), tail(e_{k+1})]$; $(v_2, v_3) \mapsto Q[e_1, e_k]$; $(v_3, v_4) \mapsto P[head(e_k), tail(e_{k+1})]$; $(v_4, v_5) \mapsto Q'[e_{k+1}, e_r]$; $(v_5, v_6) \mapsto Q'[head(e_r), head(Q')]$. The network contains Fig.1(b).

3) If $e_1, e_r \in Q'$, then the network contains Fig.1(a), which is similar to case 1).

4) If $e_1 \in Q'$ and $e_r \in Q$, then the network contains Fig.1(b), which is similar to case 2).

Likewise, if $C(s_2, t_2) \geq 2$, similar discussions can conclude that the network contains Fig.1(a) or Fig.1(c). ∎

## IV. SOLVABILITY ANALYSIS

In this section, we apply those structural results in Section III to analyze the capacity of 2-pair unicast networks. Those results deduce a complete classification of the 2-pair unicast available networks (Theorem 4.3), and an efficient algorithm to determine the solvability of a 2-pair unicast problems (Algorithm 4.5). It meanwhile provides a new proof that linear network coding is sufficient for solving the 2-pair unicast problem (Corollary 4.6). Most importantly, It is showed that the solvability of a 2-pair unicast problem is completely decided by four subsets, $\mathcal{A}_{i,j}$ for $i, j = 1, 2$ of the underlying network (Theorem 4.8).

### A. Solvability of 2-pair Unicast Problem

The results of this part are based on the technique of *informational dominance* in [8].

*Definition 4.1 ([8]):* Let $\mathcal{N} = (V, E, \{s_1, t_1\}, \{s_2, t_2\})$ be a 2-pair unicast network. We say an edge set $A$ informationally dominates an edge set $B$ if $X_B$ is a function of $X_A$ (or equivalently, $H(B|A) = 0$) for all network coding solutions, and denoted by $A \rightsquigarrow^i B$.

The informational dominance has the following properties [8]:

1) $T(i) \rightsquigarrow^i S(i)$, for $i = 1, 2$.
2) $A \rightsquigarrow^i A$, for $A \subseteq E$.
3) If $A \rightsquigarrow^i B$, and $A \rightsquigarrow^i C$, then $A \rightsquigarrow^i B \cup C$.
4) If $A \rightsquigarrow^i B$, and $B \rightsquigarrow^i C$, then $A \rightsquigarrow^i C$.
5) If $B$ is downstream of $A$, then $A \rightsquigarrow^i B$, where $B$ is *downstream* of $A$ if there is no path from $S = \{s_1, s_2\}$ to $B$ in $\mathcal{N} \setminus A$.

In the above, 1) holds by the definition of network coding solution; 2)-4) hold by the definition of informational dominance; As to 5), edge set $B$ is called *downstream* of edge set $A$ if there is no path from $S = \{s_1, s_2\}$ to $B$ in $\mathcal{N} \setminus A$, the deduced network formed by $\mathcal{N}$ deleting $A$ (see [8]), and this item holds by observing that $X_e = f_e(X_{In(e)})$ for all $e \in E$ and all the paths from $S = \{s_1, s_2\}$ to $B$ intersect $A$ (a detailed proof can be found in Lemma 11, p.2353 of [8]).

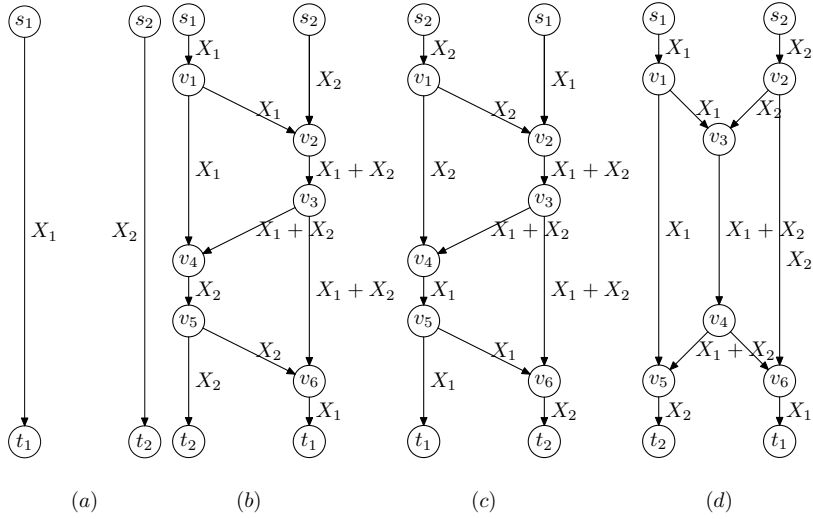

Fig. 5: Network coding solutions for Fig.1.

Given an arbitrary 2-pair unicast network $\mathcal{N} = (V, E, \{s_1, t_1\}, \{s_2, t_2\})$. If $C(s_1, t_1) \cdot C(s_2, t_2) \geq 2$, it contains a copy of Fig.1(a), Fig.1(b) or Fig.1(c). Thus $\mathcal{N}$ is available by extending the network solution of Fig.5(a), Fig.5(b), or Fig.5(c) to the whole network. That is, to transmit $X_e$ over the path $f(e)$ of $\mathcal{N}$, and not to transmit any signal over the other edges. When $C(s_1, t_1) \cdot C(s_2, t_2) = 1$ and $\mathcal{A}_{1,1} \cap \mathcal{A}_{2,2} = \emptyset$, the network contains Fig.1(a), Fig.1(b), or Fig.1(c), and then it is available. When $C(s_1, t_1) \cdot C(s_2, t_2) = 1$ and $\mathcal{A}_{1,1} \cap \mathcal{A}_{2,2} \neq \emptyset$, we have,

*Theorem 4.2:* Let $\mathcal{N} = (V, E, \{s_1, t_1\}, \{s_2, t_2\})$ be a 2-pair unicast network such that $C(s_1, t_1) \cdot C(s_2, t_2) = 1$ and $\mathcal{A}_{1,1} \cap \mathcal{A}_{2,2} \neq \emptyset$. Then $\mathcal{N}$ is available if and only if there exist an $s_1$-$t_2$ path $P_1$ and an $s_2$-$t_1$ path $P_2$ with $(P_1 \cup P_2) \cap (\mathcal{A}_{1,1} \cap \mathcal{A}_{2,2}) = \emptyset$.

*Proof:* Let $\mathcal{N}$ contain an $s_1$-$t_2$ path $P_1$ and an $s_2$-$t_1$ path $P_2$ with $(P_1 \cup P_2) \cap (\mathcal{A}_{1,1} \cap \mathcal{A}_{2,2}) = \emptyset$. By Theorem 3.9, $\mathcal{N}$ contains Fig.1(d). Then a network coding solution (shown in Fig.5(d)) can be extended to $\mathcal{N}$ (by the aforementioned manner), and the sufficiency holds.

Suppose $\mathcal{N}$ is available, and by Theorem 3.7, we take an $s_1$-$t_1$ path $P$ and an $s_2$-$t_2$ path $Q$ such that $P \cap Q = \mathcal{A}_{1,1} \cap \mathcal{A}_{2,2}$. Without loss of generality, assume that no $s_1$-$t_2$ path is disjoint with $\mathcal{A} = \mathcal{A}_{1,1} \cap \mathcal{A}_{2,2}$, we prove the result by deduce a contradiction.

Let $\mathcal{A} = \{e_1, e_2, \cdots, e_n\}$ (with the topological order) and take $e_i \in \mathcal{A}$, we claim that $T(2)$ is downstream of $\{e_i\}$. Firstly, there is no path form $s_2$ to $t_2$ in $\mathcal{N} \setminus \{e_i\}$ since $e_i \in \mathcal{A}_{2,2}$. Secondly, suppose that there exists an $s_1$-$t_2$ path $P_1$ in $\mathcal{N} \setminus \{e_i\}$, then $P_1$ intersects $\mathcal{A}$. Let $e_j \in P_1 \cap \mathcal{A}$. If $i < j$, then $P_1[s_1, e_j]$-$P[head(e_j), t_1]$ is an $s_1$-$t_1$ path without passing through $e_i \in \mathcal{A}_{1,1}$, which is a contradiction. If $i > j$, then $Q[s_2, e_j]$-$P_1[head(e_j), t_2]$ is an $s_2$-$t_2$ path without passing through $e_i \in \mathcal{A}_{2,2}$, which is again a contradiction. Hence, there is neither $s_2$-$t_2$ path nor $s_1$-$t_2$ path in $\mathcal{N} \setminus \{e_i\}$, which implies that $T(2)$ is downstream of $\{e_i\}$. Moreover, one can have that $T(1)$ is downstream of $\{e_i\} \cup S(2)$ since all $s_1$-$t_1$ paths intersect $e_i$ and all $s_2$-$t_1$ paths intersect $S(2)$.

Now we have already shown that $\{e_i\} \rightsquigarrow^i T(2)$, and $\{e_i\} \cup S(2) \rightsquigarrow^i T(1)$. Moreover, since $T(2) \rightsquigarrow^i S(2)$, one can have $\{e_i\} \rightsquigarrow^i S(2)$ by property 4). Thus $\{e_i\} \rightsquigarrow^i \{e_i\} \cup S(2) \rightsquigarrow^i T(1)$ by properties 2)-4). Using 3) again, we have $\{e_i\} \rightsquigarrow^i T(1) \cup T(2)$, which contradicts to that $e_i$ has unit capacity. The contradiction yields the necessity of the theorem. ∎

The above discussions can conclude the following theorem.

*Theorem 4.3:* The 2-pair unicast problem is solvable if and only if the underlying network contains Fig.1(a), Fig.1(b), Fig.1(c) or Fig.1(d).

*Remark 4.4:* This theorem has been independently obtained by Chih-Chun Wang and Ness B. Shroff (Theorem 3 of [19]) by using different techniques. In [19], these underlying configurations were derived based on the *path overlap conditions* (Theorem 1 of [18]), which says that a 2-pair unicast problem is solvable if and only if it satisfies some path overlap conditions. Unlike [18], [19], we formulate the network structures by *cut set (A-set) relations*. The technical differences led to different algorithms for deciding the solvability of a 2-pair unicast problem, as follows.

*Algorithm 4.5:* (Checking the solvability of a 2-pair unicast problem.)

*Input:* A 2-pair unicast network $\mathcal{N} = (V, E, \{s_1, t_1\}, \{s_2, t_2\})$.

*Output:* The solvability of the 2-pair unicast problem.

(1) : Find $C(s_1, t_1)$ and $C(s_2, t_2)$, then calculate $C(s_1, t_1) \cdot C(s_2, t_2)$.

If $C(s_1, t_1) \cdot C(s_2, t_2) = 0$, $\mathcal{N}$ is unavailable.

If $C(s_1, t_1) \cdot C(s_2, t_2) > 1$, $\mathcal{N}$ is available.

If $C(s_1, t_1) \cdot C(s_2, t_2) = 1$, goto (2).

(2) : Find $\mathcal{A}_{1,1}$ and $\mathcal{A}_{2,2}$, then calculate $\mathcal{A} = \mathcal{A}_{1,1} \cap \mathcal{A}_{2,2}$.

If $\mathcal{A}_{1,1} \cap \mathcal{A}_{2,2} = \emptyset$, $\mathcal{N}$ is available.

If $\mathcal{A}_{1,1} \cap \mathcal{A}_{2,2} \neq \emptyset$, goto (3).

(3) : Check the connectivity of $s_1$ to $t_2$ and $s_2$ to $t_1$ in $\mathcal{N}' = \mathcal{N} \setminus \mathcal{A}$.

If $C_{\mathcal{N}'}(s_1, t_2) \cdot C_{\mathcal{N}'}(s_2, t_1) = 0$, $\mathcal{N}$ is unavailable.

If $C_{\mathcal{N}'}(s_1, t_2) \cdot C_{\mathcal{N}'}(s_2, t_1) \neq 0$, $\mathcal{N}$ is available.

*End.*

In Algorithm 4.5, steps (1) and (2) can be finished in time $O(|V||E|^2)$ ([21]), and $O(|V||E|^3)$ ([20]), respectively. Step (3) can be done by a conventional breadth (or depth) first search algorithm with time $O(|V|^2)$. Note that the algorithm proposed in [18] and [19] (Corollary 1 of [18] and Corollary 1 of [19]) are based on the approach of [14] for finding $k$ edge-disjoint paths. According to [14], one need to first calculate the *levels* of all the nodes, and then use a pebbling game for the path finding process. Comparing with this approach, Algorithm 4.5 is easier to implement.

Theorem 4.3 yields the following result, which was also independently pointed out in Corollary 2 of [18] and Corollary 3 of [19].

*Corollary 4.6:* Linear network coding is sufficient to solve the 2-pair unicast problem.

## B. The 2-pair Unicast Networks with $C(s_i, t_j) = 1$

In this part, we consider the 2-pair unicast networks with $C(s_i, t_j) = 1$, for $i, j = 1, 2$.

*Lemma 4.7:* Let $\mathcal{N} = (V, E, \{s_1, t_1\}, \{s_2, t_2\})$ be a 2-pair unicast network with $C(s_i, t_j) = 1$ for $i, j = 1, 2$, and $\mathcal{A}_{1,1} \cap \mathcal{A}_{2,2} \neq \emptyset$. Then there exist an $s_1$-$t_2$ path $P_1$ and an $s_2$-$t_1$ path $P_2$ such that $(P_1 \cup P_2) \cap (\mathcal{A}_{1,1} \cap \mathcal{A}_{2,2}) = \emptyset$ if and only if $(\mathcal{A}_{1,2} \cup \mathcal{A}_{2,1}) \cap (\mathcal{A}_{1,1} \cap \mathcal{A}_{2,2}) = \emptyset$.

*Proof:* Suppose that there exist an $s_1$-$t_2$ path $P_1$ and an $s_2$-$t_1$ path $P_2$ such that $(P_1 \cup P_2) \cap (\mathcal{A}_{1,1} \cap \mathcal{A}_{2,2}) = \emptyset$. Noting that $\mathcal{A}_{1,2} \subseteq P_1$ and $\mathcal{A}_{2,1} \subseteq P_2$, we have $(\mathcal{A}_{1,2} \cup \mathcal{A}_{2,1}) \cap (\mathcal{A}_{1,1} \cap \mathcal{A}_{2,2}) = \emptyset$, which proves the necessity.

Now we prove the sufficiency. Without loss of generality, suppose all the $s_1$-$t_2$ paths intersect $\mathcal{A} = \mathcal{A}_{1,1} \cap \mathcal{A}_{2,2} = \{e_1, e_2, \cdots, e_n\} = P \cap Q$ for some $s_1$-$t_1$ path $P$ and some $s_2$-$t_2$ path $Q$, where the existence of $P$ and $Q$ is guaranteed by Theorem 3.7. Now take an arbitrary $s_1$-$t_2$ path $P_1$, and let $e_i \in P_1$ for some $1 \leq i \leq n$. One can prove that $e_j \in P_1$ for all $1 \leq j \leq n$. In fact, when $j < i$, $P_1[s_1, e_i]$-$P[head(e_i), t_1]$ is an $s_1$-$t_1$ path and hence contains $\mathcal{A}$, which implies that $e_j$ lies in $P_1$ for any $1 \leq j < i$. When $j > i$, then $Q[s_2, e_i]$-$P_1[head(e_i), t_2]$ is an $s_2$-$t_2$ path and hence contains $\mathcal{A}$. Therefore $e_j \in P_1$ for any $i < j \leq n$. The above discussions show that $\mathcal{A} \subseteq P_1$. Since $P_1$ is chosen arbitrarily, one can have that $\mathcal{A}$ is contained in all the $s_1$-$t_2$ paths, which means $\mathcal{A} \subseteq \mathcal{A}_{1,2}$, and thus $\mathcal{A}_{1,2} \cap \mathcal{A} = \mathcal{A} \neq \emptyset$. Similarly, when all $s_2$-$t_1$ paths intersect $\mathcal{A}$, we have $\mathcal{A}_{2,1} \cap \mathcal{A} = \mathcal{A} \neq \emptyset$. Therefore $(\mathcal{A}_{1,2} \cup \mathcal{A}_{2,1}) \cap (\mathcal{A}_{1,1} \cap \mathcal{A}_{2,2}) = (\mathcal{A}_{2,1} \cup \mathcal{A}_{1,2}) \cap \mathcal{A} = \mathcal{A} \neq \emptyset$, and the sufficiency holds. ∎

Now we give our main result.

*Theorem 4.8:* Let $\mathcal{N} = (V, E, \{s_1, t_1\}, \{s_2, t_2\})$ be a 2-pair unicast network such that $C(s_i, t_j) = 1$ for $i, j = 1, 2$. Then the following statements are equivalent:

(1) $\mathcal{N}$ is available.

(2) $\mathcal{N}$ contains one of the four networks depicted in Fig.1.

(3) $(\mathcal{A}_{1,2} \cup \mathcal{A}_{2,1}) \cap (\mathcal{A}_{1,1} \cap \mathcal{A}_{2,2}) = \emptyset$

*Proof:* The equivalency between (1) and (2) has already been obtained by Theorem 4.3. Also, we have shown that $\mathcal{N}$ is available if and only if $\mathcal{A}_{1,1} \cap \mathcal{A}_{2,2} = \emptyset$ or $\mathcal{A}_{1,1} \cap \mathcal{A}_{2,2} \neq \emptyset$ and there exist an $s_1$-$t_2$ path $P_1$ and an $s_2$-$t_1$ path $P_2$ such that $(P_1 \cup P_2) \cap (\mathcal{A}_{1,1} \cap \mathcal{A}_{2,2}) = \emptyset$, which is equivalent to $\mathcal{A}_{1,1} \cap \mathcal{A}_{2,2} = \emptyset$ or $\mathcal{A}_{1,1} \cap \mathcal{A}_{2,2} \neq \emptyset$ and $(\mathcal{A}_{1,2} \cup \mathcal{A}_{2,1}) \cap (\mathcal{A}_{1,1} \cap \mathcal{A}_{2,2}) = \emptyset$ by Lemma 4.7. Thus (1) and (3) are equivalent. ∎

Note that the 2-pair unicast problem just aims at supporting two unit flows. *It is adequate to assume the information edges, $S(i)$ and $T(i)$ to have unit capacities.* Under such an assumption, $\mathcal{N}$ always satisfies $C(s_i, t_j) = 1$ for $i, j = 1, 2$. Thus, the solvability of a 2-pair unicast problem is completely determined by the relations of $\mathcal{A}_{1,1}, \mathcal{A}_{2,2}, \mathcal{A}_{1,2}$, and $\mathcal{A}_{2,1}$ of the underlying network.

## V. CONCLUSIONS AND DISCUSSIONS

In this paper, we proposed a subnetwork decomposition/combination approach and decomposed a 2-pair network into four point-to-point subnetworks $\mathcal{N}_{i,j}$, for $i, j = 1, 2$. It showed that the solvability of a 2-pair unicast problem is completely determined by four link subsets, $\mathcal{A}_{1,1}, \mathcal{A}_{2,2}, \mathcal{A}_{1,2}$, and $\mathcal{A}_{2,1}$ of the underlying network. The structure of the 2-pair unicast networks was developed by analyzing the relations of the $\mathcal{A}$-sets. As a result, it deduced four specific simple available networks, such that any available 2-pair unicast network contains one copy of them and vice versa. Our results yielded an efficient algorithm to determine the solvability of the 2-pair unicast problem and a new proof that nonlinear network coding is unnecessary for solving the 2-pair unicast problem.

According to [22], the $\mathcal{A}$-set of a point-to-point network is composed by the links with *capacity rank* 1. It is reasonable to conjecture that the rate region of a general multi-source multi-sink network is merely determined by the " important links," i.e., the links with small capacity ranks. Moreover, it will be valuable to obtain an equation similar to (3) of Theorem 4.8 for the general $k$-pair unicast networks.

The four proposed underlying networks have the property that any available 2-pair unicast network contains one copy of them. From such a sense, we call them a *minimum available family under network coding* for the 2-pair unicast networks. To decide such minimum available family for 3-pair or $k$-pair unicast networks in general is still open.

We focused on directed acyclic 2-pair unicast networks in this paper. For the undirected networks, it is conjectured that network coding have no more advantages than fractional routing, which is known as the *undirected $k$-pair conjecture* [15]. To find out the *minimum available family under fractional routing* for undirected $k$-pair networks is also a more challenging topic.

## REFERENCES

[1] P. Elias, A. Feinstein, and C. E. Shannon, "Note on maximum flow through a network," *IRE Trans. Inf. Theory,* IT-2, 117-119, 1956.
[2] L.R. Ford and D. R. Fulkerson, "On the max-flow min-cut theorem of networks," in *"Linear Inequalities and Related Systems," Ann. Math. Studies,* vol. 38, Princeton, New Jersey, pp. 215-221, 1956.
[3] R.W. Yeung, "Multilevel diversity coding with distortion," *IEEE Trans. Inf. Theory,* vol. 41, no. 2, pp. 412-422, Mar. 1995.
[4] R.W. Yeung and Z. Zhang, "Distributed source coding for satellite communications," *IEEE Trans. Inf. Theory,* vol. 45, no. 3, pp. 1111-1120, May 1999.
[5] R. W. Yeung and Z. Zhang, "On symmetrical multilevel diversity coding," *IEEE Trans. Inf. Theory,* vol. 45, no. 2, pp. 609-621, Mar. 1999.
[6] R. Ahlswede, N. Cai, S.-Y. R. Li, and R. W. Yeung, "Network information flow," *IEEE Trans. Inf. Theory,* vol. 46, no. 4, pp. 1204-1216, Jul. 2000.
[7] S.-Y. R. Li, R. W. Yeung, and N. Cai, "Linear network coding," *IEEE Trans. Inf. Theory,* vol. 49, no. 2, pp. 371-381, Feb. 2003.
[8] Nicholas J.A. Harvey, Robert Kleinberg, and April Rasala Lehman, "On the capacity of information networks," *IEEE Trans. Inf. Theory,* vol. 52, pp. 2345-2364, Mar. 2006.
[9] X. Yan, J. Yang, and Z. Zhang, "An outer bound for multisource multisink network coding with minimum cost consideration,"*IEEE Trans. Inform. Theory* & *IEEE/ACM Trans. Networking,* vol. 52, no. 6, pp. 2373-2385, June 2006.
[10] G. Kramer and S. A. Savari, "Edge-cut bounds on network coding rates," *Journal of Network and Systems Management,* vol. 14, no. 1, pp. 49-67, Mar. 2006.
[11] L. Song and R. W. Yeung, "Zero-error network coding for acyclic network," *IEEE Trans. Inf. Theory,* vol. 49, no. 12, pp. 3129-3139, Dec. 2003.
[12] R. W. Yeung, "A First Course in Information Theory," New York: Kluwer/Plenum, 2002.
[13] X. Yan, R. W. Yeung, and Z. Zhang, "The capacity region for multi-source multi-sink network coding," *IEEE International Symposium on Information Theory (ISIT),* June 2007.
[14] S. Fortune, J. Hopcroft, and J. Willie, "The directed subgraph homeomorphism problem," *Theoretical Computer Science,* vol. 10. pp. 111-121, 1980.
[15] Z. Li and B. Li, "Network coding: the case of multiple unicast sessions," *The 42nd Allerton Annual Conference on Communication, Control, and Computing (Allerton 2004),* Sept. 2004.
[16] K. Jain, V. Vazirani, R. W. Yeung, and G. Yuval, "On the capacity of multiple unicast sessions in undirected graphs" *IEEE International Symposium on Information Theory (ISIT),* 2005.
[17] Nicholas J.A. Harvey, and Robert Kleinberg, "Tighter cut-based bounds for $k$-pairs communication problems," *The 43rd Annual Allerton Conference on Communication, Control, and Computing (Allerton 2005),* Sept. 2005.
[18] C.-C. Wang and N.B. Shroff, "Beyond the butterfly-a graph theoretic characterization of the feasibility of network coding with two simple unicast sessions," *IEEE International Symposium on Information Theory (ISIT),* June 2007.
[19] C.-C. Wang and N.B. Shroff, "Intersession network coding for two simple multicast sessions," *The 45th Annual Allerton Conference on Communication, Control, and Computing (Allerton 2007),* Sept. 2007.
[20] K. Cai and P. Fan, "Link decomposition of network coding," *The 12th Asia-Pacific Conference on Communications (APCC)*, Busan, Korea, Aug./Sep. 2006.
[21] J. Edmonds and R.M. Karp, "Theoretical improvements in algorithmic efficiency for network flow problems," *JACM* 19 (1972), pp. 248-264.
[22] K. Cai and P. Fan, "An Algebraic Approach to Link Failures Based on Network Coding," *IEEE Trans. Inf. Theory,* vol. 53, no. 2, pp. 775-779, Feb. 2007.